\documentclass[fleqn,usenatbib]{mnras}
\usepackage[T1]{fontenc}
\usepackage{ae,aecompl}
\usepackage{graphicx}
\usepackage{epstopdf}
\usepackage{amsmath}
\usepackage{amssymb}
\usepackage{enumerate}

\pubyear{2019}
\begin{document}

\def\func#1{\mathop{\rm #1}\nolimits}
\def\unit#1{\mathord{\thinspace\rm #1}}

\title[Modelling the light curve of iPTF16asu]{Broad-lined type Ic supernova
iPTF16asu: A challenge to all popular models}
\author[Wang et al.]{L. J. Wang,$^{1}$\thanks{%
E-mail: wanglingjun@ihep.ac.cn} X. F. Wang,$^{2}$ Z. Cano,$^{3}$\thanks{%
Juan de la Cierva Fellow.} S. Q. Wang,$^{4,5,6,7}$ L. D. Liu,$^{5,6,8}$ Z.
G. Dai,$^{5,6}$ \thanks{%
E-mail: dzg@nju.edu.cn} \newauthor J. S. Deng,$^{9,10}$ H. Yu,$^{5,6}$ B. Li,%
$^{5,1}$ L. M. Song,$^{1,11}$ Y. L. Qiu$^{9}$ and J. Y. Wei$^{9,10}$ \\
$^{1}$Astroparticle Physics, Institute of High Energy Physics, Chinese
Academy of Sciences, Beijing 100049, China \\
$^{2}$Physics Department and Tsinghua Center for Astrophysics, Tsinghua
University, Beijing 100084, China \\
$^{3}$Instituto de Astrof\'{\i}sica de Andaluc\'{\i}a (IAA-CSIC), Glorieta
de la Astronom\'{\i}a s/n, E-18008, Granada, Spain.\\
$^{4}$Guangxi Key Laboratory for Relativistic Astrophysics, School of
Physical Science and Technology, Guangxi University, Nanning 530004, China\\
$^{5}$School of Astronomy and Space Science, Nanjing University, Nanjing
210093, China \\
$^{6}$Key Laboratory of Modern Astronomy and Astrophysics (Nanjing
University), Ministry of Education, Nanjing 210093, China \\
$^{7}$Department of Astronomy, University of California, Berkeley, CA
94720-3411, USA \\
$^{8}$Department of Physics and Astronomy, University of Nevada, Las Vegas,
NV 89154, USA \\
$^{9}$Key Laboratory of Space Astronomy and Technology, National
Astronomical Observatories, Chinese Academy of Sciences, Beijing 100012,
China \\
$^{10}$School of Astronomy and Space Science, University of Chinese Academy
of Sciences, 101408 Beijing, China\\
$^{11}$University of Chinese Academy of Sciences, Chinese Academy of
Sciences, Beijing 100049, China}
\date{Accepted XXX. Received YYY; in original form ZZZ}
\maketitle

\begin{abstract}
It is well-known that ordinary supernovae (SNe) are powered by $^{56}$Ni
cascade decay. Broad-lined type Ic SNe (SNe Ic-BL) are a subclass of SNe
that are not all exclusively powered by $^{56}$Ni decay. It was suggested
that some SNe Ic-BL are powered by magnetar spin-down. iPTF16asu is a
peculiar broad-lined type Ic supernova discovered by the intermediate
Palomar Transient Factory. With a rest-frame rise time of only 4 days,
iPTF16asu challenges the existing popular models, for example, the
radioactive heating ($^{56}$Ni-only) and the magnetar+$^{56}$Ni models. Here
we show that this rapid rise could be attributed to interaction between the
SN ejecta and a pre-existing circumstellar medium ejected by the progenitor
during its final stages of evolution, while the late-time light curve can be
better explained by energy input from a rapidly spinning magnetar. This
model is a natural extension to the previous magnetar model. The mass-loss
rate of the progenitor and ejecta mass are consistent with a progenitor that
experienced a common envelope evolution in a binary. An alternative model
for the early rapid rise of the light curve is the cooling of a shock
propagating into an extended envelope of the progenitor. It is difficult at
this stage to tell which model (interaction+magnetar+$^{56}$Ni or
cooling+magnetar+$^{56}$Ni) is better for iPTF16asu. However, it is worth
noting that the inferred envelope mass in the cooling+magnetar+$^{56}$Ni is
very high.
\end{abstract}

\label{firstpage} \pagerange{\pageref{firstpage}--\pageref{lastpage}}


\section{Introduction}

Broad-lined type Ic supernovae \citep[SNe Ic-BL;][]{Woosley06} are a
particular class of stripped-envelope core collapse SNe (CCSNe) that have
received much attention since the discovery that some SNe Ic-BL are
associated with long-duration gamma-ray bursts 
\citep[LGRBs;
e.g. the GRB-SN connection][]{Woosley06, CanoWang17}.

Usually it was believed that the light curves of SNe Ic-BL, with or without
a LGRB association, are powered exclusively by $^{56}$Ni decay %
\citep[e.g.,][]{Iwamoto98,Iwamoto00,Nakamura01a}. This is verified for some
SNe, both in the early and the late phases, by measuring the Fe mass
inferred from the spectroscopic observations 
\citep[e.g.,][]{Mazzali01,
Maeda07}, although the failure of one-dimensional $^{56}$Ni model to
reproduce the late-time light curves of some SNe Ic-BL, both those with and
without an accompanying GRB stimulated the proposition of a two-component $%
^{56}$Ni model \citep{Maeda03} that mimicked the asymmetric nature of these
events \citep{Mazzali01,
Maeda03}.\ However, it appears that not all SNe Ic-BL are powered by
radioactive heating \citep{Greiner15}. Because of this fact, the magnetar
model \citep{WangHan16,
WangYu17} has been put forward in the hope to provide SNe Ic-BL with enough
kinetic energy and at the same time to power their light curves %
\citep{WangHan16, WangWang16}, based in part on the fact that the kinetic
energy of SNe Ic-BL has an upper limit that is consistent with the maximum
rotational energy of a rapidly spinning magnetar \citep{Mazzali14}. Indeed,
it was found that virtually all known SNe Ic-BL that are not associated with
LGRBs can be explained by the magnetar model \citep{WangLJCano17}.

Despite the above progress, the debate of whether the light curves of SNe
Ic-BL are powered solely by $^{56}$Ni or magnetar+$^{56}$Ni continues 
\citep{Cano16,
CanoIzzo17, Gao16, Dessart17, WangSQCano17, Chen18, Sahu18}. Here we examine
the newly discovered SN Ic-BL iPTF16asu \citep{Whitesides17} to see whether
it is consistent with either model.

iPTF16asu was discovered by the intermediate Palomar Transient Factory %
\citep[iPTF;][]{Law09, Cao16, Masci17} on 2016 May 11.26 UT. It represents
one of the most luminous type Ic SNe, with an absolute magnitude of $-20.4%
\unit{mag}$ in $g^{\prime }$ band, similar to the luminous transients %
\citep{Drout11, Drout14, Greiner15, WangWang15b, Arcavi16, Kann19}
discovered recently. Thanks to the wide field and high cadence of iPTF, the
rapid rise of the light curve of iPTF16asu (4 days to peak luminosity since
the discovery) was captured. Modelling of the light curve of iPTF16asu
indicates that iPTF16asu cannot be satisfactorily explained by just the $%
^{56}$Ni model, or the magnetar model \citep{Whitesides17}.

The structure of this paper is as follows. In Section \ref{sec:data} we
construct the bolometric light curve according to the method devised by \cite%
{Lyman14}. Our model fitting results are presented in Section \ref%
{sec:models}. We first try to separately fit the light curve of iPTF16asu
using the $^{56}$Ni-only model and magnetar model (Section \ref{sec:no-inter}%
) and confirm the finding \citep{Whitesides17} that iPTF16asu cannot be
explained solely by either one of these models by themselves. Then we fit
the light curve of iPTF16asu by including interaction (Section \ref%
{sec:inter}) between the ejecta and the circumstellar medium (CSM) or
including the cooling of a shock propagating into an extended envelope
(Section \ref{sec:cooling}). For the interaction model, we also calculate
the radio and X-ray emission (Section \ref{sec:radio-X}) so that the
predicted flux does not exceed the observational upper limits. In Section %
\ref{sec:dis} we discuss and summarise our findings.

\section{Data analysis}

\label{sec:data}

When analyzing a sample of SNe Ic-BL not associated with LGRBs, \cite%
{WangLJCano17} used the method devised by \cite{Lyman14} to construct
quasi-bolometric light curves from light curves in two individual passbands.
This method is preferred because quite often the data from near-infrared and
ultraviolet bands are not available to obtain a true bolometric light curve.
In this analysis the luminosity distances of SNe Ic-BL were calculated
according to the latest measurement of cosmological parameters: $%
H_{0}=(67.8\pm 0.9)\unit{km}\unit{s}^{-1}\unit{Mpc}^{-1}$, $\Omega
_{m}=0.308\pm 0.012$ \citep{Ade16}. \cite{WangLJCano17} also used the
photospheric velocity data obtained by \cite{Modjaz16} in a homogenous way.
Extinction and $K$ corrections were also properly taken into account %
\citep{WangLJCano17}. To make sense a direct comparison with the results of 
\cite{WangLJCano17}, in this work we have applied the same procedure to
obtain the bolometric light curve of iPTF16asu.

iPTF16asu was observed in $g^{\prime }$, $r^{\prime }$, and $i^{\prime }$
filters \citep{Whitesides17}. Some data \citep{Whitesides17} were also
acquired by \textit{Swift} UVOT \citep{Roming05} and X-ray Telescope %
\citep[XRT;][]{Burrows05}. In this paper, the bolometric light curve for the
rest-frame phases between days 1.88 to 19.49 relative to bolometric maximum
light was calculated according to color $g^{\prime }-i^{\prime }$. For
phases before day 1.88, we applied the same bolometric correction as for day
1.88 because the data for these phases are available only in the $g^{\prime }
$ band. For phases after day 19.49, we calculated the bolometric luminosity
according to the $r^{\prime }$-band magnitudes, and using the same
bolometric correction as used in the final multi-band epoch at +19.49 days.
We note that although for phases after day 19.49 data are available in $%
r^{\prime }$ and $i^{\prime }$ bands, we cannot obtain bolometric luminosity
according to these two bands because \cite{Lyman14} did not provide a method
to compute the bolometric luminosity using the $r^{\prime }-i^{\prime }$
color (see their Table 2). The data used in this paper are taken from Table
1 of \cite{Whitesides17}. These data have been corrected for foreground
extinction ($E(B-V)=0.029\unit{mag}$), arising from the sightline through
the Milky Way extinction. For the redshift $z=0.1874$ \citep{Whitesides17},
we obtained a luminosity distance $940.1\unit{Mpc}$.

The constructed bolometric light curve is presented as open circles (along
with errors) in the upper panel of Figure \ref{fig:56Ni-mag}, where the
effective temperature (color temperature; bottom left panel) and
photospheric velocity (bottom right panel) data are taken from \cite%
{Whitesides17}. We obtained a peak bolometric luminosity of $3.8\times
10^{43}\unit{erg}\unit{s}^{-1}$, which is comparable to that ($3.4\times
10^{43}\unit{erg}\unit{s}^{-1}$) given in Figure 9 of \cite{Whitesides17}.

\section{Fitting results}

\label{sec:models}

In Section \ref{sec:no-inter} we fit the observational data using the
individual $^{56}$Ni model and magnetar model. It is found that these two
popular models fail to satisfactorily reproduce the bolometric light curve
of iPTF16asu. To address this failure, in Section \ref{sec:inter} we include
energy arising from SN ejecta-CSM interaction into the models to create two-
and three-component models (CSM-interaction+$^{56}$Ni,
CSM-interaction+magnetar, and CSM-interaction+magnetar+$^{56}$Ni). The
interaction between ejecta and CSM may produce radio and X-ray emission,
which is calculated in Section \ref{sec:radio-X} to check if the predicted
flux is compatible with observational limits. Alternatively, in Section \ref%
{sec:cooling} we model the early peak of iPTF16asu by the cooling of a shock
propagating into an envelope.

\subsection{The $^{56}$Ni model and magnetar model}

\label{sec:no-inter}

In this Section we fit the bolometric light curve of iPTF16asu using the
pure-$^{56}$Ni model, pure-magnetar model and magnetar+$^{56}$Ni model. In
the pure-$^{56}$Ni model the fitting parameters are the ejecta mass $M_{%
\mathrm{ej}}$, $^{56}$Ni mass $M_{\mathrm{Ni}}$, (initial) expansion
velocity $v_{\mathrm{sc}0}$, and the gray opacity to gamma-ray photons from $%
^{56}$Ni decay $\kappa _{\gamma ,\mathrm{Ni}}$. In turn, our magnetar+$^{56}$%
Ni model includes input not only from the radioactive nickel, but also from
the spindown of a magnetar central engine. As a result, besides the above
parameters, the fitting parameters in magnetar model also include the
magnetic field strength $B_{p}$ and the initial rotation period $P_{0}$ of
the magnetar, as well as a gray opacity $\kappa _{\gamma ,\mathrm{mag}}$ to
take into account the leakage \citep{WangWang15,
Chen15} of gamma-rays from the spinning-down magnetar.

\begin{figure}
\includegraphics[width=0.5\textwidth,angle=0]{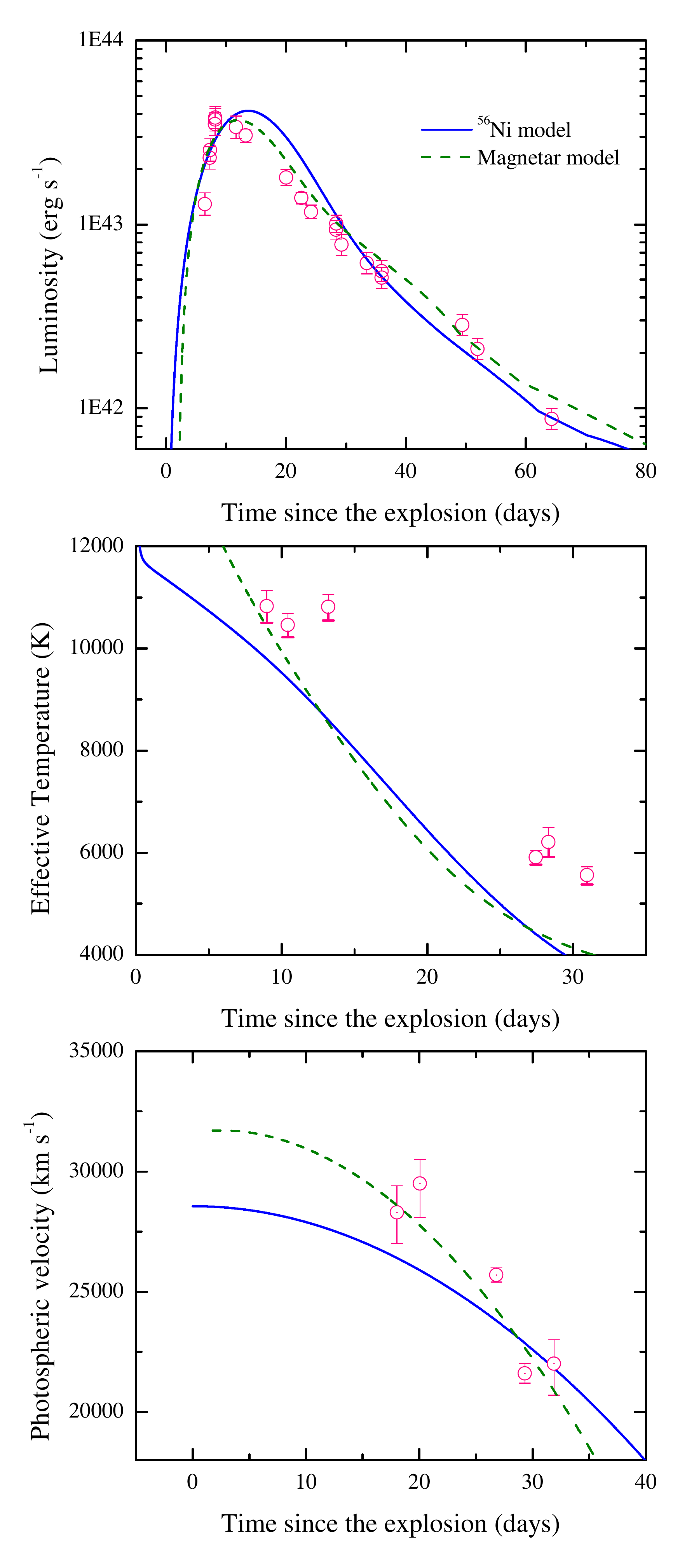}
\caption{\textit{Top}: Light curves of iPTF16asu reproduced by the $^{56}$Ni
model (solid line) and magnetar model (dashed line). \textit{Middle}:
Effective temperature in the $^{56}$Ni model (solid line) and magnetar model
(dashed line). \textit{Bottom}: Photospheric velocity in the $^{56}$Ni model
(solid line) and magnetar model (dashed line).}
\label{fig:56Ni-mag}
\end{figure}

The best-fit results are given in Figure \ref{fig:56Ni-mag} as solid lines ($%
^{56}$Ni model) and dashed lines (magnetar model), along with the fitting
parameters given in Table \ref{tbl:para-convent}. Throughout this paper we
assumed $\kappa =0.1\unit{cm}^{2}\unit{g}^{-1}$ as the fiducial optical
opacity for a type Ic SN. Here we utilise the Markov chain Monte Carlo\
(MCMC) code developed by \cite{WangYu17} to minimise the overall fitting
errors of the bolometric light curve, the color temperature and velocity
evolution. In Table \ref{tbl:para-convent} the pure-magnetar model is a
model that does not include the energy input from $^{56}$Ni. Table \ref%
{tbl:para-convent} indicates that including $^{56}$Ni does not improve the
fitting quality. Consequently, we show only the pure-magnetar fitting result
in Figure \ref{fig:56Ni-mag}.

\begin{table*}
\caption{Best-fit parameters of models without interaction.}
\label{tbl:para-convent}
\begin{center}
\begin{tabular}{lcccccccc}
\hline\hline
model & $M_{\mathrm{ej}}$ & $M_{\mathrm{Ni}}$ & $\kappa _{\gamma ,\mathrm{Ni}%
}$ & $v_{\mathrm{sc}0}$ & $T_{\mathrm{start}}$ & $B_{p}$ & $P_{0}$ & $\kappa
_{\gamma ,\mathrm{mag}}$ \\ 
\multicolumn{1}{c}{} & $\left( M_{\odot }\right) $ & $\left( M_{\odot
}\right) $ & $\left( \unit{cm}^{2}\unit{g}^{-1}\right) $ & $\left( \unit{km}%
\unit{s}^{-1}\right) $ & $\left( \unit{days}\right) $ & $\left( 10^{15}\unit{%
G}\right) $ & $\left( \unit{ms}\right) $ & $\left( \unit{cm}^{2}\unit{g}%
^{-1}\right) $ \\ \hline
pure $^{56}$Ni & $3.67_{-0.28}^{+0.29}$ & $1.76_{-0.19}^{+0.20}$ & $%
0.01_{-0.002}^{+0.003}$ & $28560_{-570}^{+670}$ & $-9.8\pm 0.3$ & NA & NA & 
NA \\ 
pure magnetar & $2.80\pm 0.11$ & NA & NA & $31700_{-670}^{+810}$ & $%
-8.0_{-0.25}^{+0.26}$ & $1.4\pm 0.03$ & $13.0\pm 0.2$ & $%
0.28_{-0.06}^{+0.13} $ \\ 
magnetar + $^{56}$Ni & $2.81_{-0.12}^{+0.16}$ & $\lesssim 0.08$ & $%
0.15_{-0.14}^{+0.34}$ & $31620_{-880}^{+850}$ & $-8.0_{-0.29}^{+0.28}$ & $%
1.4_{-0.04}^{+0.05}$ & $13.1\pm 0.3$ & $0.25_{-0.06}^{+0.11}$ \\ \hline
\end{tabular}%
\end{center}
\par
\textbf{Notes.} In these fits we fixed $\kappa =0.1\unit{cm}^{2}\unit{g}%
^{-1} $. NA means not applicable.
\end{table*}

As can been seen from Figure \ref{fig:56Ni-mag}, neither the $^{56}$Ni model
nor the magnetar model can account for the rapid rise in the light curve. In
the pure-$^{56}$Ni model, the ratio $M_{\mathrm{Ni}}/M_{\mathrm{ej}%
}=1.76/3.67=0.48$ is unrealistic and significantly larger than the
theoretical upper limit of 0.2 \citep{Umeda08}. We conclude that iPTF16asu
cannot be explained by the three models considered here (radioactivity, a
magnetar, or the combined model), confirming the conclusion of \cite%
{Whitesides17}.

\subsection{Models including interaction}

\label{sec:inter}

Inspecting the light curve of iPTF16asu indicates that the most striking
feature is its rapid rise to peak luminosity. It is just this rapid rise
that cannot be fitted by any of the models. For an SN that reaches a peak
luminosity ($3.8\times 10^{43}\unit{erg}\unit{s}^{-1}$) comparable to that
of superluminous SNe \citep[SLSNe; ][]{GalYam12, Gal-Yam18, Moriya18, Wang19}%
, a rapidly rotating magnetar 
\citep{Kasen10, Woosley10, Chatzopoulos12, Inserra13, Nicholl14,
Metzger15, WangWang15, WangWang16, Dai16, LiuWangSQ17, Yu17} or interaction
between the ejecta and CSM surrounding the progenitor 
\citep{Chatzopoulos12, Chatzopoulos13, Ginzburg12,
Nicholl14, Chen15} has been proposed as energy source. We show in Figure \ref%
{fig:56Ni-mag} that the magnetar model cannot capture the rapid rise of the
light curve. Ejecta-CSM interaction, however, can reproduce a rapidly rising
light curve, as demonstrated recently by \cite{LiuWangLJ18}, who applied a
multiple-interaction model to reproduce the undulating light curves of
iPTF13dcc \citep{Vreeswijk17}\ and iPTF15esb \citep{Yan17}. The spectra of
iPTF16asu and iPTF13dcc \citep[Figure 5 in][]{Vreeswijk17} near peak are
pretty featureless, consistent with CSM-ejecta SNe. However, the spectra of
iPTF15esb \citep[Figure 1 in][]{Yan17} near peak light show prominent
absorption at rest-frame $\sim 4100\unit{%
\text{\AA}%
}$.

Without the corresponding narrow Balmer (IIn) or He lines, the early spectra
of iPTF16asu \citep{Whitesides17} somewhat resemble the spectra of SNe IIn %
\citep[e.g.,][]{Andrews17, Nyholm17}, which are generally believed to
originate from the interaction between the ejecta and CSM. The spectra 2-3
weeks later have good resemblance to SNe Ic-BL; see Figure 12 in \cite%
{Whitesides17} which shows iPTF16asu relative to the prototypical SN Ic-BL
1998bw. The lack of narrow emission lines is not an argument against the
ejecta-CSM interaction because there are a variety of reasons for the
suppressing of emission lines 
\citep[e.g.,][]{Chevalier11, Moriya12,
Chatzopoulos13}. If the interaction model is correct for iPTF16asu, the CSM
interacting with the ejecta should be ejected recently during the final
evolution of the progenitor (see Section \ref{sec:dis}). This implies that
the CSM was quite possibly fast-moving. In this case the ionising radiation
from the forward shock may only give rise to intermediate-width emission
lines. Furthermore, the CSM interacting with iPTF16asu's ejecta is quite
dense and the ionising radiation was not enough to ionise the dilute
material beyond the CSM shell. It is beyond the scope of this work, and
indeed the limits of the analytical models employed in our investigation, to
predict the spectral behaviour and evolution of iPTF16asu, which should be
done with more sophisticated modelling. Instead, the focus of this work is
to reproduce its photometric behaviour, of which its physical processes are
inferred from.

Although a multiple-interaction model is certainly plausible to reproduce
the bolometric light curve of iPTF16asu, in this paper we pursue an
interaction+magnetar+$^{56}$Ni model, hereafter referred to as
interaction-plus-magnetar model\footnote{%
For a luminous SN, because the contribution to the bolometric luminosity
from $^{56}$Ni is usually much smaller than that from a magnetar, to save
words, we call it an interaction-plus-magnetar model and omit $^{56}$Ni when
saying a model actually including energy input from CSM-interaction, a
magnetar, and $^{56}$Ni.}. We propose that the rapidly rising peak is caused
by ejecta-CSM interaction, while the later-time slow decay is powered by a
magnetar and $^{56}$Ni. Such an interaction+magnetar+$^{56}$Ni model was
applied to fit the light curve of iPTF13ehe \citep{WangLiu16}, who
constructed a model in which the main peak is powered by magnetar spin-down
while the late-time excess luminosity is caused by an interaction between
ejecta and CSM.

We chose to examine the interaction+magnetar+$^{56}$Ni model rather than
multiple-interaction model for iPTF16asu because it was shown that a large
sample of SNe Ic-BL are consistent with the magnetar+$^{56}$Ni model %
\citep{WangLJCano17}. The inclusion of interaction serves as a natural
extension to the magnetar+$^{56}$Ni model.

For the interaction between the ejecta and the CSM, we adopt the model
developed by \cite{Chevalier82} and \cite{Chevalier94}\footnote{%
We found some mistakes in \cite{Chatzopoulos12,Chatzopoulos13}. The correct
expressions are presented in Appendix \ref{app:ejecta-csm}.}. Because we
assume that the light curve peak is caused by an interaction, the CSM is
very close to the SN progenitor. We assume that the CSM has a density
profile of a stellar wind, that is, the power-law index for CSM density
profile is $s=2$. The density profile index of the inner ejecta is set $%
\delta =0$, while the slope of the outer ejecta is set $n=7$. The
dimensionless radius of break in the SN ejecta density profile from the
inner component to the outer component is set $x_{0}=0.3$.

\begin{figure}
\includegraphics[width=0.5\textwidth,angle=0]{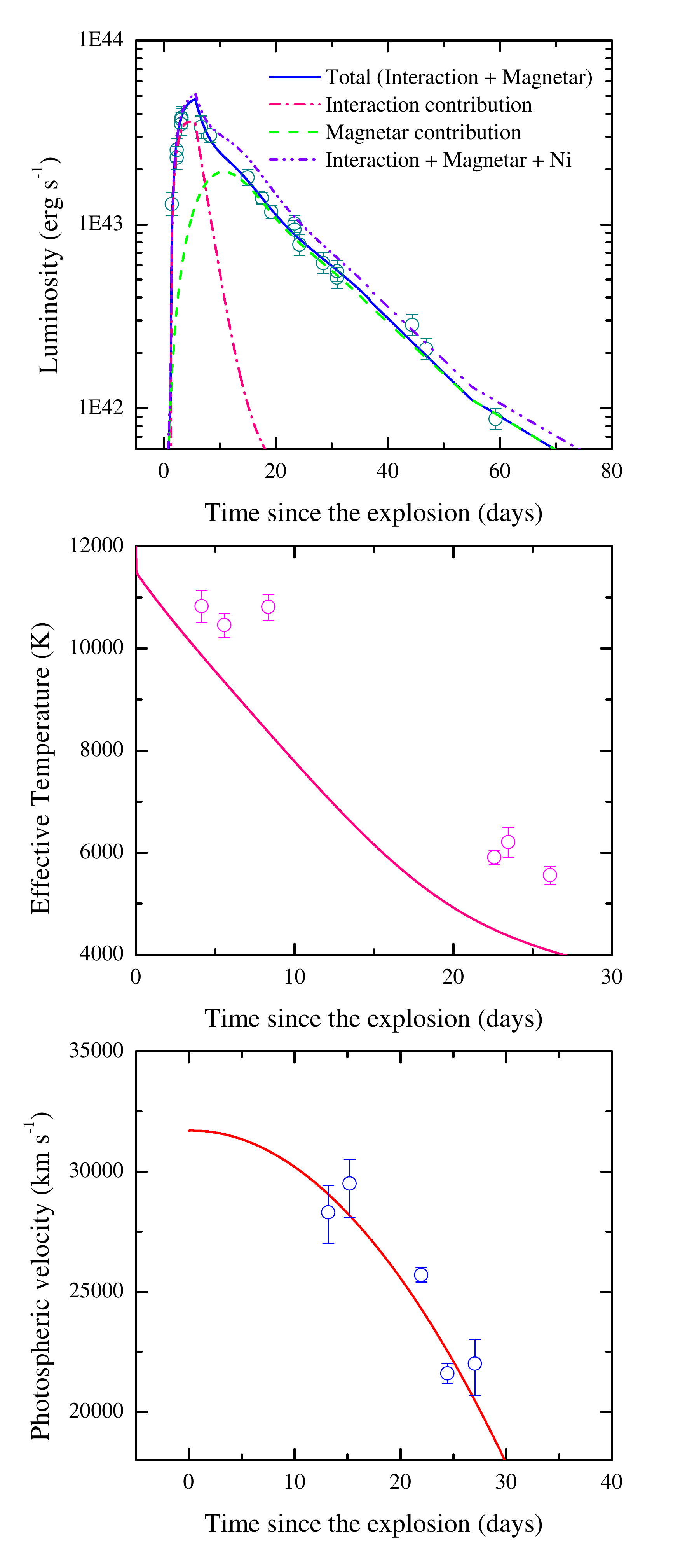}
\caption{\textit{Top}: Light curve of iPTF16asu reproduced by the
interaction-plus-magnetar model. The pink dot-dashed line is the
contribution of the interaction, the green dashed line is the contribution
of the magnetar, while the solid blue line is their total contribution. The
dot-dot-dashed violet line is the light curve including $0.2M_{\odot }$ of $%
^{56}$Ni. Note that this curve only serves as an indication that no more
than $0.2M_{\odot }$ of $^{56}$Ni is allowed by the light curve. \textit{%
Middle}: Effective temperature. \textit{Bottom}: Photospheric velocity.}
\label{fig:mag-inter}
\end{figure}

The fitting results of the interaction+magnetar model are shown in Figure %
\ref{fig:mag-inter} with best-fit parameters listed in Table \ref%
{tbl:para-inter}, where $M_{\mathrm{ej}}$ is the ejecta mass of the SN, $M_{%
\mathrm{CSM}}$ is the CSM mass, $R_{\mathrm{CSM},\mathrm{in}}$ is the inner
radius of CSM, $\rho _{\mathrm{CSM},\mathrm{in}}$ is the density of CSM at
radius $R_{\mathrm{CSM},\mathrm{in}}$, $\epsilon $ is the radiation
efficiency. Other parameters have the same meaning as in the magnetar model.
In this model, the effective ejecta mass for the photons emanated from the
magnetar to diffuse is $M_{\mathrm{tot}}=M_{\mathrm{ej}}+M_{\mathrm{CSM}}$.
This approximation is valid because the CSM is very close to the SN ejecta
so that ejecta and CSM can be treated as a continuous mass distribution.

\begin{table*}
\caption{Best-fit parameters of models with interaction.}
\label{tbl:para-inter}
\begin{center}
\begin{tabular}{lcccccccccc}
\hline\hline
model & $M_{\mathrm{ej}}$ & $M_{\mathrm{CSM}}$ & $R_{\mathrm{CSM,in}}$ & $%
\rho _{\mathrm{CSM,in}}$ & $\epsilon $ & $v_{\mathrm{sc}0}$ & $T_{\mathrm{%
start}}$ & $B_{p}$ & $P_{0}$ & $\kappa _{\gamma ,\mathrm{mag}}$ \\ 
\multicolumn{1}{c}{} & $\left( M_{\odot }\right) $ & $\left( M_{\odot
}\right) $ & $\left( 10^{15}\unit{cm}\right) $ & $\left( 10^{-13}\unit{g}%
\unit{cm}^{-3}\right) $ &  & $\left( \unit{km}\unit{s}^{-1}\right) $ & $%
\left( \unit{days}\right) $ & $\left( 10^{15}\unit{G}\right) $ & $\left( 
\unit{ms}\right) $ & $\left( \unit{cm}^{2}\unit{g}^{-1}\right) $ \\ \hline
interaction + magnetar & $1.56$ & $0.61$ & $0.33$ & $9$ & $0.06$ & $31700$ & 
$-4.7$ & $1.5$ & $19$ & $0.28$ \\ \hline\hline
& $M_{\mathrm{ej}}$ & $M_{\mathrm{CSM}}$ & $R_{\mathrm{CSM,in}}$ & $\rho _{%
\mathrm{CSM,in}}$ & $\epsilon $ & $v_{\mathrm{sc}0}$ & $T_{\mathrm{start}}$
& $M_{\mathrm{Ni}}$ & $\kappa _{\gamma ,\mathrm{Ni}}$ &  \\ 
& $\left( M_{\odot }\right) $ & $\left( M_{\odot }\right) $ & $\left( 10^{15}%
\unit{cm}\right) $ & $\left( 10^{-13}\unit{g}\unit{cm}^{-3}\right) $ &  & $%
\left( \unit{km}\unit{s}^{-1}\right) $ & $\left( \unit{days}\right) $ & $%
\left( M_{\odot }\right) $ & $\left( \unit{cm}^{2}\unit{g}^{-1}\right) $ & 
\\ \hline
interaction + $^{56}$Ni & $1.8$ & $0.64$ & $0.33$ & $9$ & $0.07$ & $29700$ & 
$-4.9$ & $0.5$ & $0.057$ &  \\ \hline
\end{tabular}%
\end{center}
\par
\textbf{Notes.} In these fits we fixed $\kappa =0.1\unit{cm}^{2}\unit{g}%
^{-1} $.
\end{table*}

Figure \ref{fig:mag-inter} indicates that the interaction+magnetar model
provides a fairly good fitting to the light curve, temperature evolution and
velocity evolution. In this model, $M_{\mathrm{tot}}$ is determined by the
velocity evolution curve, while the small value of $M_{\mathrm{CSM}}$ is
required by the rapid rise of the light curve before peak luminosity. Other
parameters, e.g., $B_{p}$, $P_{0}$, $\kappa _{\gamma ,\mathrm{mag}}$ are
typical for the sample of SNe Ic-BL \citep{WangLJCano17}.

To accurately determine the amount of $^{56}$Ni mass required by the light
curve, observational data up to at least 100 days are necessary.
Nevertheless, we show in Section \ref{fig:56Ni-mag} that for a magnetar+$%
^{56}$Ni model not involving interaction, the needed $^{56}$Ni mass is very
small. Here we do not apply a MCMC code to constrain the $^{56}$Ni mass. In
Figure \ref{fig:mag-inter} we draw a curve including $0.2M_{\odot }$ of $%
^{56}$Ni, in addition to the contribution of the magnetar and the
interaction. From Figure \ref{fig:mag-inter} we see that this model does not
also allow for $^{56}$Ni mass much larger than $0.2M_{\odot }$, i.e., the
maximum amount of $^{56}$Ni that can be synthesised by a spinning-down
magnetar \citep{Nishimura15,
Suwa15}. This indicates that the interaction+magnetar model is a reasonable
model for iPTF16asu.

\begin{figure}
\includegraphics[width=0.5\textwidth,angle=0]{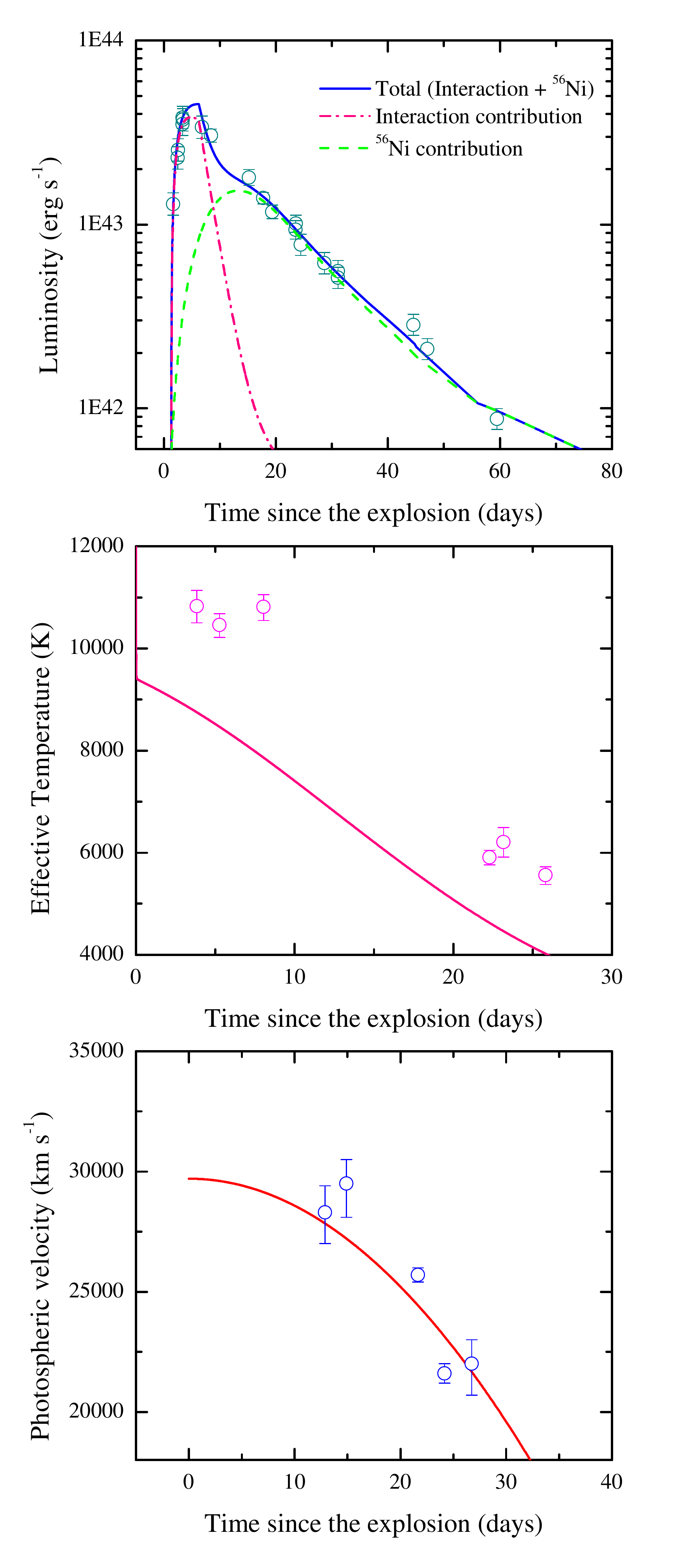}
\caption{\textit{Top}: Light curve of iPTF16asu reproduced by the
interaction plus $^{56}$Ni model. The pink dot-dashed line is the
contribution of interaction, the green dashed line is the contribution of $%
^{56}$Ni, while the solid blue line is their total contribution. \textit{%
Middle}: Effective temperature. \textit{Bottom}: Photospheric velocity.}
\label{fig:56Ni-inter}
\end{figure}

As a comparison, we would like to examine whether an interaction+$^{56}$Ni
model is consistent with iPTF16asu. We show the fitting results of such a
model in Figure \ref{fig:56Ni-inter} along with the best-fitting parameters
listed in Table \ref{tbl:para-inter}. It can be seen from Figure \ref%
{fig:56Ni-inter} that although the fitting quality of the interaction+$^{56}$%
Ni model is reasonably good, the fitting parameters are of concern. The
nickel mass needed to explain the large luminosity of iPTF16asu implies an
unphysical nickel mass--ejecta mass ratio of 0.28, which is larger than the
theoretical maximum value of $\sim 0.2$ \citep{Umeda08}. The required $^{56}$%
Ni mass is also atypically large among CCSNe. This indicates that iPTF16asu
was unlikely to be primarily powered by radioactive decay of $^{56}$Ni.

\subsection{Radio and X-ray emission in the ejecta-CSM interaction model}

\label{sec:radio-X}

The forward shock and reverse shock induced by the ejecta-CSM collision
accelerate the particles behind their respective shock front to relativistic
speed. In the meantime, magnetic field is also generated by these two
shocks. The motion of relativistic electrons in the magnetic field generates
synchrotron emission, which, in the SN environment, is long-lived radio
emission \citep[see e.g.,][]{Chevalier17, Chandra18}.

It is usually assumed that the magnetic energy density is a constant
fraction of the internal energy density of the shocked gas, i.e., $%
B^{2}/8\pi =\epsilon _{B}\rho _{\mathrm{CSM}}V_{s}^{2}$, where $\rho _{%
\mathrm{CSM}}$ is the circumstellar density, $V_{s}$ is the shock velocity
relative to the CSM, $\epsilon _{B}<1$ is a constant. For the relativistic
electrons, there are two possible assumptions. One is that a fixed fraction
of the shocked electrons are accelerated, i.e., $U_{\mathrm{rel}}\propto
\rho _{\mathrm{CSM}}$. An alternative assumption is that the energy density
of relativistic electrons is proportional to the swept up thermal energy,
i.e., $U_{\mathrm{rel}}\propto \rho _{\mathrm{CSM}}V_{s}^{2}$ %
\citep[e.g.,][]{Chevalier17}. Here we assume $U_{\mathrm{rel}}=\epsilon
_{e}\rho _{\mathrm{CSM}}V_{s}^{2}$ \citep{Fransson98}, where $\epsilon
_{e}<1 $ is another constant.

For a stellar wind CSM, the forward shock moves at velocity 
\citep[see
e.g.,][]{Chevalier17, Chandra18}%
\begin{equation}
V_{s}=\frac{dR_{s}}{dt}=\frac{n-3}{n-2}V_{\mathrm{ej}},
\end{equation}%
where $V_{\mathrm{ej}}$ is the ejecta velocity, and $n$ is the power-law
density index of the outer part of the ejecta (see Appendix), while the
reverse shock moves at velocity%
\begin{equation}
V_{\mathrm{rev}}=V_{\mathrm{ej}}-V_{\mathrm{s}}=\frac{V_{\mathrm{ej}}}{n-2}
\end{equation}%
relative to the ejecta.

The relativistic electrons suffer from several cooling processes. First of
all, the electrons may lose energy by synchrotron emission, with a cooling
time%
\begin{equation}
t_{\mathrm{syn}}=\frac{6\pi m_{e}c}{\sigma _{T}\gamma B^{2}},
\end{equation}%
where $m_{e}$ is the electron mass, $\gamma $ is the Lorentz factor of the
electron emitting synchrotron photons at frequency%
\begin{equation}
\nu _{c}=\frac{eB\gamma ^{2}}{m_{e}c},
\end{equation}%
where $e$ is the electron charge.

The electrons may also lose energy by inverse Compton scattering, with a
cooling time%
\begin{equation}
t_{\mathrm{Comp}}=\frac{3m_{e}c}{4\sigma _{T}\gamma U_{\mathrm{rad}}},
\end{equation}%
where $U_{\mathrm{rad}}$ is the energy density of photons. For an SN, $U_{%
\mathrm{rad}}$ is dominated by the thermal emission at the photosphere,
i.e., $U_{\mathrm{rad}}=L/4\pi R^{2}c$, where $L$ is the bolometric
luminosity of the SN. The scattering of optical photons at the SN
photosphere by relativistic electrons produces X-ray emission.

The Coulomb interaction between electrons and other charged particles is
another mechanism to lose energy \citep{Fransson98}. Putting all these
together, the accelerated electrons cool within a timescale %
\citep{Fransson98}%
\begin{equation}
t_{\mathrm{loss}}=\left( \frac{1}{t_{\mathrm{syn}}}+\frac{1}{t_{\mathrm{Comp}%
}}+\frac{1}{t_{\mathrm{Coul}}}+\frac{1}{t}\right) ^{-1},
\end{equation}%
where the last term takes adiabatic losses into account.

The electrons are accelerated into a power-law spectrum by the shock, $%
N\left( E\right) dE=N_{0}E^{-p}dE$, where $N_{0}$ is a constant that is
given by%
\begin{equation}
N_{0}=\left( p-2\right) U_{e}E_{\min }^{p-2}.
\end{equation}%
Here $E_{\min }\approx m_{e}c^{2}$ is the minimum energy of the accelerated
electrons. The radio flux density at frequency $\nu $ is given by %
\citep{Chandra18}%
\begin{eqnarray}
F_{\nu }^{\mathrm{SSA}} &=&\frac{\pi R^{2}}{D^{2}}\frac{c_{5}}{c_{6}}%
B^{-1/2}\left( \frac{\nu }{2c_{1}}\right) ^{5/2}[1-\exp (-\tau _{\nu }^{%
\mathrm{SSA}})], \\
\tau _{\nu }^{\mathrm{SSA}} &=&\int\limits_{0}^{s}\kappa (\nu )ds\approx
\kappa (\nu )s=\left( \frac{\nu }{2c_{1}}\right) ^{-\frac{p+4}{2}}\left( 
\frac{4}{3}fRc_{6}N_{0}B^{\frac{p+2}{2}}\right) ,  \label{eq:tau_SSA}
\end{eqnarray}%
where $D$ is the distance of the SN to observer, $c_{1}$, $c_{5}$, $c_{6}$
are constants defined in \cite{Pacholczyk70}, $f\approx 3/4$ is a filling
factor of the emitting material. Here it is assumed that the absorption is
synchrotron self-absorption dominated, for which the optical depth is given
by $\left( \ref{eq:tau_SSA}\right) $.

The X-ray emission produced by inverse Compton scattering is%
\begin{equation}
\frac{dE_{\mathrm{IC}}}{dt}=\frac{4}{3}\sigma _{T}c\gamma ^{2}\beta ^{2}U_{%
\mathrm{rad}},
\end{equation}%
where $\beta \approx 1$ is the dimensionless velocity of the emitting
photons. The Lorentz factor of the emitting electrons is%
\begin{equation}
\gamma \approx \left( \frac{3}{4}\frac{\epsilon _{1}}{\epsilon }\right)
^{1/2},
\end{equation}%
where $\epsilon $ and $\epsilon _{1}$ are the energies of the seed photons
and scattered photons, respectively. The seed photons are assumed to be
blackbody photons from the SN photosphere, and $\epsilon =3.6kT_{\mathrm{bb}%
} $ \citep{Felten66}, where $T_{\mathrm{bb}}$ is the blackbody temperature
of the SN emission.

With the above consideration, we can calculate the radio and X-ray emission
produced by the ejecta-CSM interaction. Radio observations of iPTF16asu were
carried out on 2016 June 13 and 2017 January 10 and a $3\sigma $ limit of $%
\approx 17\unit{\mu Jy}$ at $6.2\unit{GHz}$ for both epochs was derived %
\citep{Whitesides17}. These two epochs correspond to rest-frame 28.2 and
205.9 days after explosion, respectively. In addition, X-ray observations
were carried out at phases 7.4, 13.4, and 19.2 days after explosion, with
flux limits in the energy band $0.3-10\unit{keV}$ of $2.5\times 10^{43}\unit{%
erg}\unit{s}^{-1}$, $1.1\times 10^{43}\unit{erg}\unit{s}^{-1}$, and $%
1.5\times 10^{43}\unit{erg}\unit{s}^{-1}$, respectively \citep{Whitesides17}%
. These three epochs correspond to rest-frame 6.2, 11.3, 16.2 days after
explosion.

We estimate the radio flux density and X-ray emission by assuming the
following typical parameters: $p=3$ \citep{Chevalier98}, $\epsilon
_{e}=3\times 10^{-3}$, $\epsilon _{B}=10^{-3}$. The calculation gives radio
flux density $0.05\unit{\mu Jy}$, and $15.5\unit{\mu Jy}$ at rest-frame
epochs 28.2 and 205.9 days after explosion, respectively. The X-ray flux at
rest-frame epochs 6.2, 11.3, and 16.2 days are $3.0\times 10^{40}\unit{erg}%
\unit{s}^{-1}$, $1.4\times 10^{40}\unit{erg}\unit{s}^{-1}$, $7.8\times
10^{39}\unit{erg}\unit{s}^{-1}$, respectively. The values are all consistent
with the observational upper limits.

\subsection{Magnetar model including shock cooling}

\label{sec:cooling}

An alternative view is that the rapid rise of the light curve of iPTF16asu
is caused by the cooling of a shock propagating into the extended envelope
of the progenitor, as suggested by \cite{Whitesides17}. We adopt the model
developed by \cite{Piro15} for shock cooling. We attribute the late-time
light curve of iPTF16asu to the energy injection of a magnetar because the
radioactive decay of $^{56}$Ni is not enough to power such high luminosity.

\begin{table*}
\caption{Best-fit parameters of models with shock cooling.}
\label{tbl:para-cooling}
\begin{center}
\begin{tabular}{cccccccc}
\hline\hline
$M_{\mathrm{ej}}$ & $M_{\mathrm{env}}$ & $R_{\mathrm{env}}$ & $v_{\mathrm{sc}%
0}$ & $T_{\mathrm{start}}$ & $B_{p}$ & $P_{0}$ & $\kappa _{\gamma ,\mathrm{%
mag}}$ \\ 
$\left( M_{\odot }\right) $ & $\left( M_{\odot }\right) $ & $\left( 10^{11}%
\unit{cm}\right) $ & $\left( \unit{km}\unit{s}^{-1}\right) $ & $\left( \unit{%
days}\right) $ & $\left( 10^{15}\unit{G}\right) $ & $\left( \unit{ms}\right) 
$ & $\left( \unit{cm}^{2}\unit{g}^{-1}\right) $ \\ \hline
$1.6\pm 0.3$ & $1.0_{-0.2}^{+0.3}$ & $37.4\pm 9$ & $35500_{-8300}^{+15000}$
& $-3.8\pm 0.1$ & $1.5\pm 0.05$ & $18\pm 0.2$ & $0.18_{-0.03}^{+0.04}$ \\ 
\hline
\end{tabular}%
\end{center}
\par
\textbf{Notes.} In this fit we fixed $\kappa =0.1\unit{cm}^{2}\unit{g}^{-1}$.
\end{table*}

To estimate the uncertainties of the fitting parameters, we extended the
code \citep{WangYu17} to take into account the shock-cooling. The fitting
results of this cooling+magnetar+$^{56}$Ni model are presented in Figure \ref%
{fig:mag-cooling} with best-fit parameters listed in Table \ref%
{tbl:para-cooling}. Comparison of Tables \ref{tbl:para-inter} and \ref%
{tbl:para-cooling} shows that the magnetar-related parameters are quite
similar. The total mass $M_{\mathrm{tot}}=M_{\mathrm{ej}}+M_{\mathrm{env}}$
is also similar to the total mass in the magnetar+interaction model because $%
M_{\mathrm{tot}}$ determines the effective diffusion timescale.

\begin{figure}
\includegraphics[width=0.5\textwidth,angle=0]{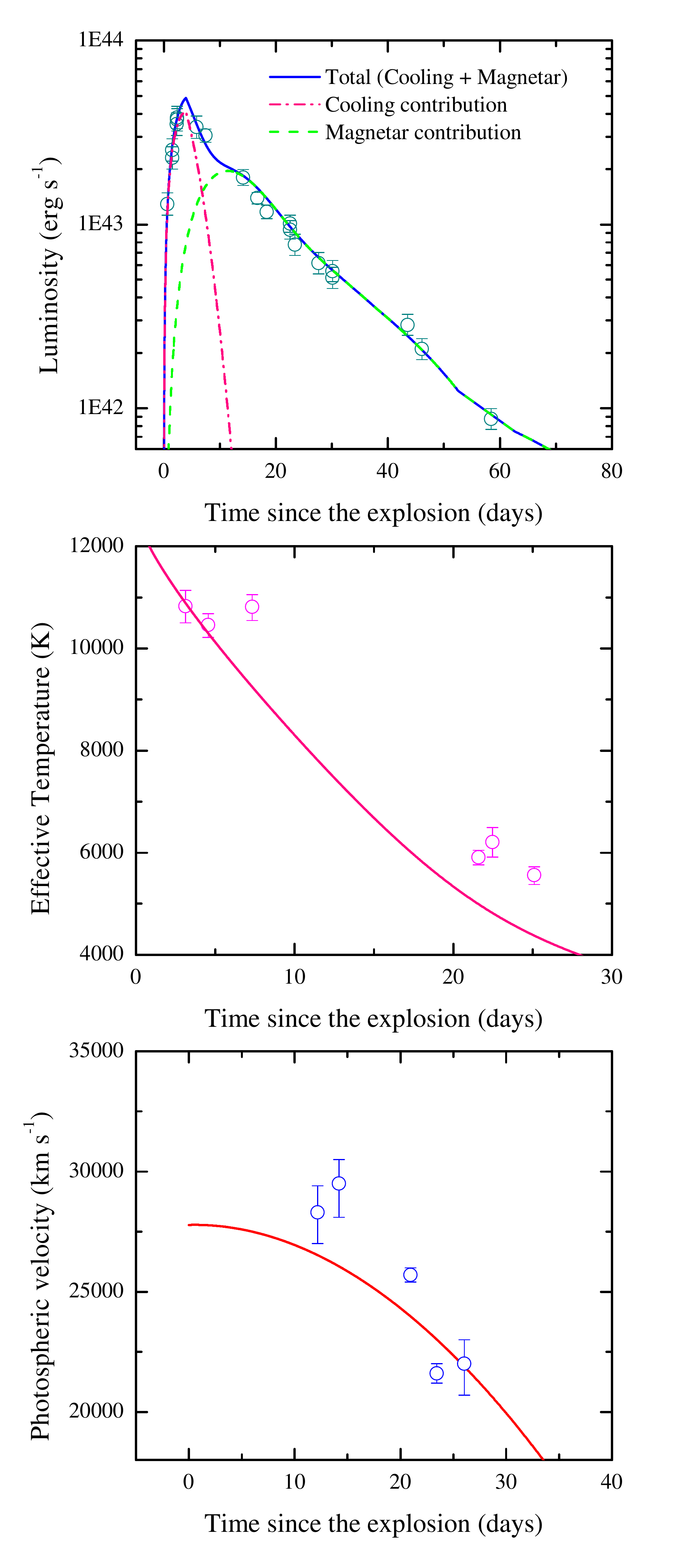}
\caption{\textit{Top}: Light curve of iPTF16asu reproduced by the shock
cooling plus magnetar model. The pink dot-dashed line is the contribution of
shock cooling, the green dashed line is the contribution of magnetar, while
the solid blue line is their total contribution. \textit{Middle}: Effective
temperature. \textit{Bottom}: Photospheric velocity.}
\label{fig:mag-cooling}
\end{figure}

In this model, the ejecta expanding at an initial velocity $\sim 36000\unit{%
km}\unit{s}^{-1}$ were slowed down rapidly when propagating into the
extended envelope. The envelope has a radius $\sim 2.6\times 10^{12}\unit{cm}%
=37R_{\odot }$, typical for a type Ic progenitor. The relatively massive
envelope mass $\left( M_{\mathrm{env}}=1.0M_{\odot }\right) $ is necessary
to produce the high peak luminosity of iPTF16asu. The ejecta mass $\left( M_{%
\mathrm{ej}}=1.6M_{\odot }\right) $ of iPTF16asu is also typical among type
Ic SNe.

\section{Discussion and Conclusions}

\label{sec:dis}

SNe Ic-BL are a subclass of SNe Ic with very diverse properties. Some SNe
Ic-BL are of normal luminosity, while others are as luminous as
superluminous supernovae. The discovery of iPTF16asu adds further diversity
to SNe Ic-BL and poses some challenges to current models of SNe Ic-BL. By
introducing an interaction between ejecta and CSM, or a shock cooling, the
light curve of iPTF16asu can be reasonably reproduced by the magnetar+$^{56}$%
Ni model. We show that even including an interaction or shock cooling, the $%
^{56}$Ni-only model is unlikely to give an acceptable interpretation to
iPTF16asu. Nevertheless, because of the short observational time of this SN,
a firm conclusion cannot be drawn at this stage. There is some evidence of
aspherical explosions of SNe Ic-BL 
\citep{Patat01, Mazzali01,Mazzali05,
Maeda08, Taubenberger09, Milisavljevic15}.\footnote{%
However, it was found that the double-peaked oxygen lines, which are
interpreted as off-axis views of a GRB jet or unipolar blob ejections, are
not rare and may not indicate aspherical explosions \citep{Modjaz08}.} The
slow material in aspherical explosions can only reveal themselves when
late-time light curve is available.

The interaction model \citep{Chevalier82, Chatzopoulos13} assumed a fixed
photospheric radius, while the collision between the ejecta and the CSM
piles up the CSM in front of the ejecta and eventually pushes the CSM to
move as a whole. It is therefore necessary to check whether the assumption
of fixed photospheric radius is valid during the phase when the contribution
from the interaction dominates over that from the magnetar. The parameters
listed in Table \ref{tbl:para-inter} imply a CSM with outer radius $R_{%
\mathrm{CSM},\mathrm{out}}=1.3\times 10^{15}\unit{cm}$, which means that the
CSM will be crossed by the ejecta in 4 days. This timescale is slightly
shorter than but comparable to the rise time of the light curve. Therefore
the assumption of a fixed photospheric radius is approximately valid.

After the CSM is crossed, the photosphere will expand and recede as in the
usual magnetar model \citep{WangWang16}. Therefore strictly speaking, the
photospheric velocity should experience an abrupt increase at $\sim 4\unit{%
days}$ after the explosion. However, because the first observational
velocity data point is at a time $\sim 13\unit{days}$ after the explosion,
we do not complicate the model to include this abrupt increase in the
velocity evolution.

The CSM properties allow us to infer the mass-loss history of iPTF16asu. The
mass-loss rate is $4\pi R_{\mathrm{CSM},\mathrm{in}}^{2}\rho _{\mathrm{CSM},%
\mathrm{in}}v_{l}$, where $v_{l}$ is the mass-loss velocity. Substituting in
the parameters in Table \ref{tbl:para-inter} yields a mass-loss rate $%
0.2M_{\odot }\unit{yr}^{-1}\left( v_{l}/100\unit{km}\unit{s}^{-1}\right) $.%
\footnote{%
Such mass-loss rate is comparable with that of some of the most famous SNe
II, for example, SN 1994W \citep[$\sim0.2M_{\odot}%
\unit{yr}^{-1}$,][]{Chugai04} and SN 1995G \citep[$\sim0.1M_{\odot}%
\unit{yr}^{-1}$,][]{Chugai03}.} The typical wind velocity of an SN Ic-BL is $%
100-3000\unit{km}\unit{s}^{-1}$ \citep{Smith14, Margutti17}. If the CSM were
a stellar wind, we get a mass-loss rate $\left( 0.2-6.6\right) M_{\odot }%
\unit{yr}^{-1}$. This is inconsistent with the properties of SNe Ic-BL,
which have typical mass-loss rate $\sim 10^{-6}M_{\odot }\unit{yr}^{-1}$ %
\citep{Smith14}. If the CSM is ejected during the common envelope (CE)
evolution, for the typical ejection velocity $10\unit{km}\unit{s}^{-1}$ %
\citep{Smith14} we get a mass-loss rate $0.02M_{\odot }\unit{yr}^{-1}$,
which is consistent with CE ejection. The inner and outer radii of the CSM
indicate that the CSM was ejected during a phase $\sim 11-44\unit{yr}$
before the SN explosion.

The CSM is also consistent with luminous blue variable (LBV) eruptions which
have mass-loss rates $10^{-2}-1M_{\odot }\unit{yr}^{-1}$ and ejection
velocity $100-6000\unit{km}\unit{s}^{-1}$ \citep{Margutti17}. It is argued
that LBVs are unstable massive stars with initial mass $20M_{\odot }\lesssim
M_{\mathrm{ini}}\lesssim 25M_{\odot }$ \citep{Groh13a, Groh13b}. Such
massive single stars are unlikely to have ejected mass as low as $%
1.6M_{\odot }$ (see Table \ref{tbl:para-inter}) when they explode as SNe.

Theoretical modelling indicates that the progenitor of an SN Ic (broad-lined
or not) could be a low-mass star in a binary or a massive single star (%
\citealt{Nomoto94, Nomoto95}, see \citealt{Smartt09} for recent review). The
ejecta mass of iPTF16asu $M_{\mathrm{ej}}$ is compatible with a binary
origin \citep{Yoon15, Fremling16}. This is consistent with the above
inference that iPTF16asu has experienced with a CE evolution.

For the interaction+magnetar model, Table \ref{tbl:para-inter} indicates an
initial explosion energy $9.4\times 10^{51}\unit{erg}$, which is
significantly larger than what can be provided by neutrino heating %
\citep{Janka16, Bollig17}. The required energy may be provided by jets %
\citep{Soker16, Soker17}.

Apart from the interaction+magnetar model, the shock-cooling+magnetar model
works equally well for iPTF16asu. At present the data do not favor one model
over the other. However, the envelope mass in this model is very high, $\sim
1M_{\odot }$. Future observations of iPTF16asu-like SNe Ic-BL could shed
more light on the evolution and progenitors of SNe Ic-BL.

\section*{Acknowledgements}

We thank David Alexander Kann, Noam Soker and the anonymous referee for
helpful comments of the manuscript. This work is supported by the National
Program on Key Research and Development Project of China (Grant No.
2016YFA0400801 and 2017YFA0402600), and the National Natural Science
Foundation of China (Grant Nos. 11573014, 11533033, 11673006 and 11833003).
X. Wang is supported by the National Natural Science Foundation of China
(NSFC grants 11325313 and 11633002), and the National Program on Key
Research and Development Project (grant no. 2016YFA0400803). S.Q.W. and
L.D.L. are also supported by China Scholarship Program to conduct research
at U.C. Berkeley and UNLV, respectively.

\appendix

\section{Ejecta-CSM interaction model}

\label{app:ejecta-csm}

The SN ejecta enter into a homologous expansion stage after a few expansion
timescales $R_{p}/v_{\mathrm{SN}}$, where $R_{p}$ is the progenitor radius,
and $v_{\mathrm{SN}}$ is the expansion velocity of the SN. The outer density
profile of the ejecta can be reasonably described by a steep power law with $%
n>5$ \citep{Chevalier82}%
\begin{equation}
\rho _{\mathrm{ej},\mathrm{outer}}=t^{-3}\left( \frac{r}{tg}\right)
^{-n}=t^{-3}\left( \frac{v}{g}\right) ^{-n},  \label{eq:ejecta-profile-outer}
\end{equation}%
while the inner part can be well described by a flat profile%
\begin{equation}
\rho _{\mathrm{ej},\mathrm{inner}}=t^{-3}\left( \frac{r}{tf}\right)
^{-\delta }=t^{-3}\left( \frac{v}{f}\right) ^{-\delta },
\label{eq:ejecta-profile-inner}
\end{equation}%
where $\delta <3$. The continuity condition $\rho _{\mathrm{ej},\mathrm{inner%
}}\left( v_{t}\right) =\rho _{\mathrm{ej},\mathrm{outer}}\left( v_{t}\right) 
$ at the transition velocity $v_{t}$ yields the following relation%
\begin{equation}
f=v_{t}\left( \frac{g}{v_{t}}\right) ^{n/\delta }.
\end{equation}%
$v_{t}$ and $g$ can be determined by the SN explosion energy $E_{\mathrm{SN}%
} $ and its ejecta mass $M_{\mathrm{ej}}$ \citep{Chevalier94}%
\begin{eqnarray}
v_{t} &=&\left[ \frac{2\left( 5-\delta \right) \left( n-5\right) E_{\mathrm{%
SN}}}{\left( 3-\delta \right) \left( n-3\right) M_{\mathrm{ej}}}\right]
^{1/2}, \\
g^{n} &=&\frac{1}{4\pi \left( n-\delta \right) }\frac{\left[ 2\left(
5-\delta \right) \left( n-5\right) E_{\mathrm{SN}}\right] ^{\left(
n-3\right) /2}}{\left[ \left( 3-\delta \right) \left( n-3\right) M_{\mathrm{%
ej}}\right] ^{\left( n-5\right) /2}}.
\end{eqnarray}%
However, $v_{t}$ is unobservable, and a constant $x_{0}=v_{t}/v_{\mathrm{SN}%
} $ is introduced \citep{Chatzopoulos12}. The explosion energy is therefore%
\begin{equation}
E_{\mathrm{SN}}=\frac{\left( 3-\delta \right) \left( n-3\right) }{2\left(
5-\delta \right) \left( n-5\right) }M_{\mathrm{ej}}\left( x_{0}v_{\mathrm{SN}%
}\right) ^{2}.
\end{equation}

It is assumed that there exists one or several CSMs, whose density profile
satisfies the following power law%
\begin{equation}
\rho _{\mathrm{CSM}}=qr^{-s},
\end{equation}%
where $s<3$. The normalisation factor $q$ is set to be $q=\rho _{\mathrm{CSM}%
,1}r_{1}^{s}$, where $\rho _{\mathrm{CSM},1}$ is the density at radius $%
r_{1} $.

The CSM is swept up by the expansion of the SN ejecta. The interaction
between the ejecta and CSM triggers the formation of forward shock into CSM
and reverse shock back into the ejecta. It is assumed that part of the shock
energy is converted into radiation 
\begin{equation}
L=\epsilon \frac{dE}{dt}=\epsilon \frac{d}{dt}\left( \frac{1}{2}M_{\mathrm{sw%
}}v_{\mathrm{sh}}^{2}\right) =\epsilon \left( M_{\mathrm{sw}}v_{\mathrm{sh}}%
\dot{v}_{\mathrm{sh}}+\frac{1}{2}\dot{M}_{\mathrm{sw}}v_{\mathrm{sh}%
}^{2}\right) ,  \label{eq:shock-lum}
\end{equation}%
where $\epsilon $ is the conversion efficiency of the shock energy into
radiation, $M_{\mathrm{sw}}$ is the swept-up mass, and $v_{\mathrm{sh}}$ is
the shock velocity.

The dynamics of the forward shock and reverse shock are described by a
self-similar solution \citep{Chevalier82}%
\begin{eqnarray}
R_{F}(t) &=&R_{\mathrm{CSM},\mathrm{in}}+\beta _{F}\left( \frac{Ag^{n}}{q}%
\right) ^{1/\left( n-s\right) }t^{\left( n-3\right) /\left( n-s\right) },
\label{eq:radius-forward-shock} \\
R_{R}(t) &=&R_{\mathrm{CSM},\mathrm{in}}+\beta _{R}\left( \frac{Ag^{n}}{q}%
\right) ^{1/\left( n-s\right) }t^{\left( n-3\right) /\left( n-s\right) },
\label{eq:radius-reverse-shock}
\end{eqnarray}%
where $R_{\mathrm{CSM},\mathrm{in}}$ is the inner radius of the CSM, $\beta
_{F}$ and $\beta _{R}$ are two constants given in Table 1 of \cite%
{Chevalier82} as $R_{1}/R_{c}$ and $R_{2}/R_{c}$, respectively. It is easy
to find that the CSM mass swept-up by the forward shock is %
\citep{Chatzopoulos12}%
\begin{eqnarray}
M_{\mathrm{sw},F} &=&4\pi \int_{R_{\mathrm{CSM}}}^{R_{F}\left( t\right)
}\rho _{\mathrm{CSM}}\left( r\right) r^{2}dr  \notag \\
&=&\frac{4\pi \beta _{F}^{3-s}}{3-s}q^{\left( n-3\right) /\left( n-s\right)
}\left( Ag^{n}\right) ^{\left( 3-s\right) /\left( n-s\right) }t^{\left(
n-3\right) \left( 3-s\right) /\left( n-s\right) }.
\label{eq:mass-forward-shock}
\end{eqnarray}%
Combined with Equation $\left( \ref{eq:radius-forward-shock}\right) $, one
finds%
\begin{equation}
L_{F}(t)=\frac{2\pi \epsilon }{(n-s)^{3}}\left( Ag^{n}\right) ^{\frac{5-s}{%
n-s}}q^{\frac{n-5}{n-s}}(n-3)^{2}(n-5)\beta _{F}^{5-s}t^{\frac{2n+6s-ns-15}{%
n-s}},
\end{equation}%
for $t_{i}<t<t_{i}+t_{\mathrm{FS},\mathrm{BO}}$, where $t_{i}\simeq R_{%
\mathrm{CSM},\mathrm{in}}/v_{\mathrm{SN}}$, $t_{\mathrm{FS},\mathrm{BO}}$ is
the break-out time of the forward shock.

Similarly, the ejecta mass swept up by the reverse shock is%
\begin{eqnarray}
M_{\mathrm{sw},R} &=&4\pi \int_{R_{R}\left( t\right) }^{R_{\mathrm{SN}%
}\left( t\right) }\rho _{\mathrm{ej}}\left( r\right) r^{2}dr  \notag \\
&=&\frac{4\pi g^{n}}{\left( n-3\right) v_{\mathrm{SN}}^{n-3}}\left\{ \left[ 
\frac{v_{\mathrm{SN}}}{\beta _{R}}\left( \frac{t^{3-s}q}{Ag^{n}}\right)
^{1/\left( n-s\right) }\right] ^{n-3}-1\right\} .
\end{eqnarray}%
In the above equation, we set $\rho _{\mathrm{ej}}=\rho _{\mathrm{ej},%
\mathrm{outer}}$, while $R_{\mathrm{SN}}\left( t\right) =v_{\mathrm{SN}}t$
is the ejecta radius. By setting $\rho _{\mathrm{ej}}=\rho _{\mathrm{ej},%
\mathrm{outer}}$ we actually ignore the transition of the ejecta density
from the outer steep profile to the inner shallow profile. We decide to do
so because the self-similar solution $\left( \ref{eq:radius-reverse-shock}%
\right) $ is valid only for steep ejecta profile \citep{Chevalier82}.

To find the luminosity of the reverse shock, we have to evaluate the
quantities in the comoving frame of the shock front of the homologously
expanding ejecta. In this frame the reverse shock moves at velocity 
\begin{equation}
\tilde{v}_{R}=\dot{R}_{R}-\frac{R_{R}}{t}=-\frac{3-s}{n-s}\beta _{R}\left( 
\frac{Ag^{n}}{q}\right) ^{1/\left( n-s\right) }t^{-\left( 3-s\right) /\left(
n-s\right) }.
\end{equation}%
Substituting this quantity and its derivative into Equation $\left( \ref%
{eq:shock-lum}\right) $ yields%
\begin{equation}
L_{R}=\frac{2\pi \epsilon g^{n}}{\beta _{R}^{n-5}}\frac{n-5}{n-3}\left( 
\frac{3-s}{n-s}\right) ^{3}\left( \frac{q}{Ag^{n}}\right) ^{\frac{n-5}{n-s}%
}t^{\frac{2n+6s-ns-15}{n-s}}
\end{equation}%
for $t_{i}<t<t_{i}+t_{\mathrm{RS},\ast }$, where $t_{\mathrm{RS},\ast }$ is
the termination time of the reverse shock.

Assuming the shocks are far from the photosphere, one can obtain the SN
luminosity by substituting the following energy input term%
\begin{equation}
L_{\mathrm{inp}}(t)=L_{F}(t)+L_{R}(t)
\end{equation}%
into the Arnett's equation \citep{Arnett82}.

The break-out time of the forward shock can be determined by equating
Equation $\left( \ref{eq:mass-forward-shock}\right) $ to the swept-up mass
of the CSM when shock break out occurs%
\begin{equation}
t_{\mathrm{FS},\mathrm{BO}}=\left[ \frac{\left( 3-s\right) M_{\mathrm{CSM}}}{%
4\pi \beta _{F}^{3-s}q^{\left( n-3\right) /\left( n-s\right) }\left(
Ag^{n}\right) ^{\left( 3-s\right) /\left( n-s\right) }}\right] ^{\left(
n-s\right) /\left( n-3\right) \left( 3-s\right) }.
\end{equation}%
The reverse shock termination time can be similarly found%
\begin{equation}
t_{\mathrm{RS},\ast }=\left\{ \frac{\beta _{R}}{v_{\mathrm{SN}}}\left( \frac{%
Ag^{n}}{q}\right) ^{1/\left( n-s\right) }\left[ \frac{\left( n-3\right) v_{%
\mathrm{SN}}^{n-3}M_{\mathrm{ej}}}{4\pi g^{n}}+1\right] ^{1/\left(
n-3\right) }\right\} ^{\left( n-s\right) /\left( 3-s\right) }.
\end{equation}

Comparing the above equations with those given by \cite{Chatzopoulos12}
shows that the expressions of \cite{Chatzopoulos12} for the forward shock
are correct. There are some mistakes in Equation $\left( B5\right) $ of \cite%
{Chatzopoulos12}; there should be an extra factor $\beta _{R}^{5-n}\left(
n-5\right) /\left( n-3\right) $ in Equation $\left( B7\right) $ of \cite%
{Chatzopoulos12}, but Equation $\left( B11\right) $ in \cite{Chatzopoulos12}
is correct. When calculating the output light curve, there should be an
extra factor $\left( n-5\right) /\left( n-3\right) $ in the reverse shock
terms of Equations $\left( 20\right) $ and $\left( 21\right) $ in \cite%
{Chatzopoulos12}. With this extra factor, the luminosities from forward
shock and reverse shock are zero in the case of $n=5$.

\bsp
\label{lastpage}

\end{document}